\documentclass{article}

\usepackage{PRIMEarxiv}

\usepackage[utf8]{inputenc} 
\usepackage[T1]{fontenc}    
\usepackage{hyperref}       
\usepackage{url}            
\usepackage{booktabs}       
\usepackage{amsfonts}       
\usepackage{nicefrac}       
\usepackage{microtype}      
\usepackage{lipsum}
\usepackage{xcolor} 
\usepackage{fancyhdr}       
\usepackage{graphicx}       
\usepackage{amsmath}
\usepackage{float}
\usepackage{amsthm}
\usepackage{amssymb}
\graphicspath{{media/}}     

\title{Hierarchy of extreme-event predictability in turbulence revealed by machine learning}

\author{
  Yuxuan Yang \\
  National University of Singapore \\
  Singapore \\
  \texttt{y.yang.65@u.nus.edu} \\
    \And
  Chenyu Dong \\
  National University of Singapore \\
  Singapore \\
  \texttt{chenyu.dong@u.nus.edu} \\
   \And
  Gianmarco Mengaldo \\
  National University of Singapore \\
  Singapore \\
  \texttt{mpegim@nus.edu.sg} \\
}

\begin{document}
\maketitle

\begin{abstract}
Extreme-event predictability in turbulence is strongly state dependent, yet event-by-event predictability horizons are difficult to quantify without access to governing equations or costly perturbation ensembles.
Here we train an autoregressive conditional diffusion model on direct numerical simulations of the two-dimensional Kolmogorov flow and use a CRPS-based skill score to define an event-wise predictability horizon. 
Enstrophy extremes exhibit a pronounced hierarchy: forecast skill persists from $\approx 1$ to $> 4$ Lyapunov times across events. 
Spectral filtering shows that these horizons are controlled predominantly by large-scale structures. 
Extremes are preceded by intense strain cores organizing quadrupolar vortex packets, whose lifetime sharply separates long- from short-horizon events. 
These results identify coherent-structure persistence as a governing mechanism for the predictability of turbulence extremes and provide a data-driven route to diagnose predictability limits from observations.
\end{abstract}


\section{Introduction}\label{sec:introduction}

Predicting the future evolution of natural phenomena such as the Earth’s atmosphere is central to science and carries major societal and economic implications~\cite{zhu2002economic}. 
Such systems are high-dimensional and multiscale (often with intrinsic or effective stochasticity), and their chaotic dynamics limits predictability to a finite horizon~\cite{eckmann1985ergodic,pikovsky2016lyapunov,Risken2012,Pavliotis2014,lorenz1963deterministic,siegert2016prediction}. These limits are most consequential for rare, high-impact excursions far from the normal system's behavior -- extreme events -- where early warning is essential~\cite{de2013predictability,yuan2024limits}.

Recent numerical studies indicate that extreme-event predictability is strongly \emph{event dependent}: different extremes can exhibit markedly different predictability~\cite{vela2024large}.
This variability suggests that extremes are not universally less predictable and may be controlled by identifiable physical mechanisms~\cite{lucarini2023typicality}.
However, the origins of this event-to-event spread in predictability remain poorly understood. 
For instance, in the context of fluid mechanics, it is unclear what flow organizations (if any) separate more predictable extremes from rapidly unpredictable ones.

A key open question is whether such differences can be diagnosed \emph{directly from data}, without access to the governing equations.
Addressing this question is nontrivial because extremes are rare and therefore difficult to learn and characterize reliably with data-driven models~\cite{olivetti2024data,wei2025xai4extremes}.
Nevertheless, equation-free predictability diagnostics are highly desirable, both for realistic systems where only observations are available and for building mechanistic understanding that can inform forecasting and control~\cite{vonich2024predictability}.

To place this question in context, it is useful to recall how predictability is traditionally quantified in dynamical systems.
The evaluation of predictability traces back to the sensitivity to initial conditions introduced by Lorenz~\cite{lorenz1963deterministic}, later formalized through the largest Lyapunov exponent~\cite{oseledec1968multiplicative}, which characterizes the exponential growth of infinitesimal perturbations.
Subsequently, various extensions have been developed to quantify predictability over finite temporal scales, thereby enabling the investigation of local predictability in nonlinear regimes~\cite{eckhardt1993local,boffetta1998extension,aurell1997predictability,ding2007nonlinear,huai2017quantifying}.
In practice, many of these approaches are inherently ensemble-dependent, where multiple trajectories are generated by perturbing the initial conditions of a known forward operator~\cite{vela2024predictability}.
Such requirements pose significant challenges: the governing equations are often unknown in realistic settings, the computational cost of ensembles can be prohibitive, and results frequently exhibit high sensitivity to the choice of initial perturbations~\cite{vela2024large}.
Data-driven approaches offer equation-free alternatives that bypass the need for an explicit forward operator. 
These methods include local dynamical indices rooted in extreme value theory~\cite{lucarini2016extremes,dong2025spatio,fang2025dynamical}, recent state-dependent predictability metrics such as time-lagged recurrence~\cite{dong2025time}, and information-theoretic measures that quantify predictability through divergences between forecast and climatological distributions~\cite{delsole2004predictability,schneider1999conceptual,kleeman2002measuring,cover1999elements}. 
These approaches characterize predictability as a property of the system or of typical states, but not as a forecast problem conditioned on a particular event. 
As a result, they do not provide an event-wise score curve from which a concrete event-wise predictability horizon (also referred to as limit) can be extracted.

Recent advances in machine learning provide a paradigm shift, whereby data-driven approaches can directly learn the evolution of forecast distributions~\cite{kohl2026benchmarking,shokar2024stochastic}, effectively acting as surrogate models. 
This new route enables the explicit estimation of event-conditioned predictability limits, bringing ensemble-style forecast verification into an equation-free, data-driven setting.

In this work, we introduce a diffusion-based ensemble forecasting framework that implements this principle and allows quantifying event-wise predictability limits of extreme events from data. 
Forecast distributions are generated by an autoregressive conditional diffusion model~\cite{kohl2026benchmarking}, enabling stable probabilistic predictions at long lead times. 
These forecast distributions are evaluated using probabilistic scoring rules, and an event-wise predictability horizon is defined from the lead-time decay of forecast skill, expressed in units of the Lyapunov time.

Applying the proposed framework to a well-known two-dimensional turbulence testbed, the Kolmogorov flow, we find a pronounced hierarchy of predictability across extremes, with predictability horizons spanning from $\sim 1$ to $>4$ Lyapunov times. 
The resulting separation into high- and low-predictability events is consistent with rankings obtained from DNS perturbation ensembles, supporting the physical relevance of the data-driven estimates. 
Scale-filtering further shows that these horizons are controlled predominantly by large-scale modes. 
Finally, coherent-structure analysis identifies the precursor organization responsible for this hierarchy: extreme events are anchored to intense strain cores that organize quadrupolar vortices, and the temporal persistence of these quadrupoles distinguishes highly predictable extremes from those with short predictability. 
Together, these findings identify coherent-structure persistence as the mechanism setting extreme-event predictability horizons and show that event-wise, skill-based diagnostics provide a practical route to quantify these predictability limits directly from data.

\section{Methodology}\label{sec:methodology}
We consider a homogeneous and isotropic turbulent flow contained in a doubly periodic square domain of area $L^2 = (2\pi)^2$, driven by a sinusoidal body force with a specified wavenumber.
This system is known as the Kolmogorov flow~\cite{fylladitakis2018kolmogorov}, and it is governed by the incompressible Navier-Stokes equations
\begin{subequations}\label{eq:NS}
\begin{align}
\nabla\cdot \mathbf{u} &= 0, \label{eq:1}\\
\frac{\partial \mathbf{u}}{\partial t} + \mathbf{u}\cdot\nabla \mathbf{u} &= -\nabla p + \nu \nabla^2 \mathbf{u} + \mathbf{f},\label{eq:2}
\end{align}
\end{subequations}
where $\mathbf{u}$ is the \emph{velocity field}, $p$ is the \emph{kinematic pressure}, $\nu$ is the \emph{kinematic viscosity}, and $\mathbf{f}$ is the forcing term defined as
\begin{equation}
\mathbf{f}(y) = \bigg\{f_0\bigg(\frac{L}{2\pi}\bigg)\cos\!\bigg(\frac{2\pi n_f}{L}\,y\bigg), \; 0 \bigg\},\label{eq:3}
\end{equation}
with forcing amplitude $f_0$ and integer forcing mode $n_f=4$ (corresponding to physical wavenumber $k_f=2\pi n_f/L$). 
The forcing term in Eq.~\eqref{eq:3} only acts on the $x$-direction. 
Following Refs.~\cite{chandler2013invariant,farazmand2017variational}, we define the Reynolds number as
\begin{equation}
Re = \frac{\sqrt{f_0}\,(L/2\pi)^2}{\nu},\label{eq:4}
\end{equation} 
and set $Re = 100$, a regime known to exhibit extreme events. 
The two-dimensional incompressible Navier–Stokes equations~\eqref{eq:NS} are solved by direct numerical simulation (DNS) on a $128\times128$ grid in a doubly periodic domain of size $L^2=(2\pi)^2$.
Extreme events in this flow configuration take the form of intermittent bursts of energy dissipation, and are identified as sudden surges in the instantaneous spatially averaged enstrophy $\Omega = \langle\omega\rangle^2$.
Specifically, we define extreme events as those states with enstrophy $\Omega$ belonging to the top $1\%$ of enstrophy distribution in the dataset.
This criterion corresponds to a threshold value of approximately $\Omega_{\mathrm{th}} \approx 8.5$.
Using the deterministic DNS dataset created on the $128 \times 128$ squared domain of size $L^2=(2\pi)^2$, we construct a diffusion-based autoregressive probabilistic model that learns the conditional distribution of future velocity fields given the current state, i.e., $p(x_{t+1}\,|\,x_t)$~\cite{ho2020denoising,kohl2026benchmarking}. 
The model follows an autoregressive denoising diffusion formulation trained on velocity-field evolution, implemented with a U-Net backbone~\cite{ronneberger2015u} and various established smaller architecture modernizations~\cite{dhariwal2021diffusion}.
Although the training data consist of deterministic realizations, the diffusion model learns the conditional distribution of future states given the present state, thereby providing a stochastic representation of the flow dynamics.
Ensemble forecasts are then generated by sampling from the learned distribution, without explicitly introducing perturbations to the initial conditions. 
Once trained, the model recursively samples future states, enabling probabilistic forecasts over arbitrarily long horizons and facilitating event-wise predictability assessment at extended lead times. More details regarding the model are provided in the Supplemental Material.

Forecast quality is assessed with the continuous ranked probability score ($\mathrm{CRPS}$)~\cite{gneiting2007strictly,anderson1996method}.
For an extreme event $e$ whose enstrophy attains its maximum at time $t_e$, we assess the forecast for the scalar observable $\Omega(t_e)$ using predictions initialized at time $t = t_e - T$, i.e., at lead time $T = t_e - t$ (in Sec.~\ref{sec:results}, we equivalently write $T = -(t - t_e)$ to represent time relative to the event peak).
We evaluate forecast skill at the enstrophy peak (in analogy with Ref.~\cite{vela2024large}) rather than at the threshold-crossing time because the latter depends on the arbitrary choice of the detection threshold and shifts when that threshold is varied.
In contrast, the peak time $t_e$ is uniquely defined, ensuring that the resulting predictability estimates are robust to the threshold used to identify extremes.  

The $\mathrm{CRPS}$ for event $e$ at lead time $T$ is given by
\begin{equation}
    \mathrm{CRPS}_e^T(D,\Omega_{e}) = \int_\mathbb{R}(F_D(\Omega)-H(\Omega-\Omega_e))^2 \mathrm{d}\Omega
    \label{eq:crps}
\end{equation}
where $D$ is the distribution of the forecasted enstrophy, $F_D$ is the cumulative distribution function of the forecasted enstrophy, $H$ is the unit step function, and $\Omega_e \in \mathbb{R}$ is the ground truth of enstrophy at the event peak. 
A vanishing $\mathrm{CRPS}^T_e$ in Eq.~\eqref{eq:crps} corresponds to a perfect probabilistic forecast, whereas large values indicate degraded predictive performance due to either substantial bias between the predicted mean and the ground truth or excessive forecast spread.

To determine whether a forecast at lead time $T$ retains skill for predicting extreme events, we compare $\mathrm{CRPS}_e^T$ against a reference $\mathrm{CRPS}_{\mathrm{ref}}$ defined as
\begin{equation}
    \mathrm{CRPS}_{\mathrm{ref}} = \int_\mathbb{R}(F_{L}(\Omega)-H(\Omega-\Omega_{th}))^2\mathrm{d}\Omega
    \label{eq:ref-crps}
\end{equation}
where $F_L$ is the cumulative distribution function of the long-term distribution of enstrophy denoted by subscript $L$, and $\Omega_{th}$ is the threshold of extreme events.
The reference score defined in equation~\eqref{eq:ref-crps} represents the best performance attainable under climatological prediction, since using the extreme-event threshold yields the lower bound of the CRPS in the absence of state-dependent information.

Using Eqs.~\eqref{eq:crps} and ~\eqref{eq:ref-crps}, we introduce the event-wise predictability score for event $e$ at lead time $T$
\begin{equation}
    {S}_e(T) = 1-\frac{\mathrm{CRPS}_e^T}{\mathrm{CRPS}_{\mathrm{ref}}}.
    \label{eq:predictability}
\end{equation}
The event-wise predictability score ${S}_e(T)$ has an upper bound of one, with ${S}_e(T) = 1$ corresponding to a perfect forecast. 
Values ${S}_e(T) \leq 0$ indicate that the forecast skill in predicting extreme events is no better than that obtained from the long-term distribution (i.e., climatology).
The event-wise predictability score ${S}_e(T)$ measures forecast skill relative to a reference prediction, in the spirit of the continuous ranked probability skill score (CRPSS)~\cite{WILKS2011301}.
The key difference here is that ${S}_e(T)$ is computed for individual extreme events and specific lead times, rather than as an average over an entire dataset. 
This allows us to directly assess the predictability of particular extreme events, enabling further event-wise analysis.

The predictability horizon (also referred to as limit) $T^*_e$ for event $e$ is then the largest lead time $T$ such that ${S}_e(T)>0$; equivalently, it is the lead time at which the forecast $\mathrm{CRPS}_e^T$ degrades to the long-term reference $\mathrm{CRPS}_{\mathrm{ref}}$.
In practice, we evaluate ${S}_e(T)$ for each detected extreme event using different lead times.
The statistics of the resulting predictability limits $T^*_e$ are expressed in both units of the system's Lyapunov time $T_\lambda = 3.6\sqrt{f_{0}}$~\cite{vela2024large} and simulation time units.

\section{Results}\label{sec:results}

\begin{figure}[H]
\centering
\includegraphics[width=0.8\textwidth]{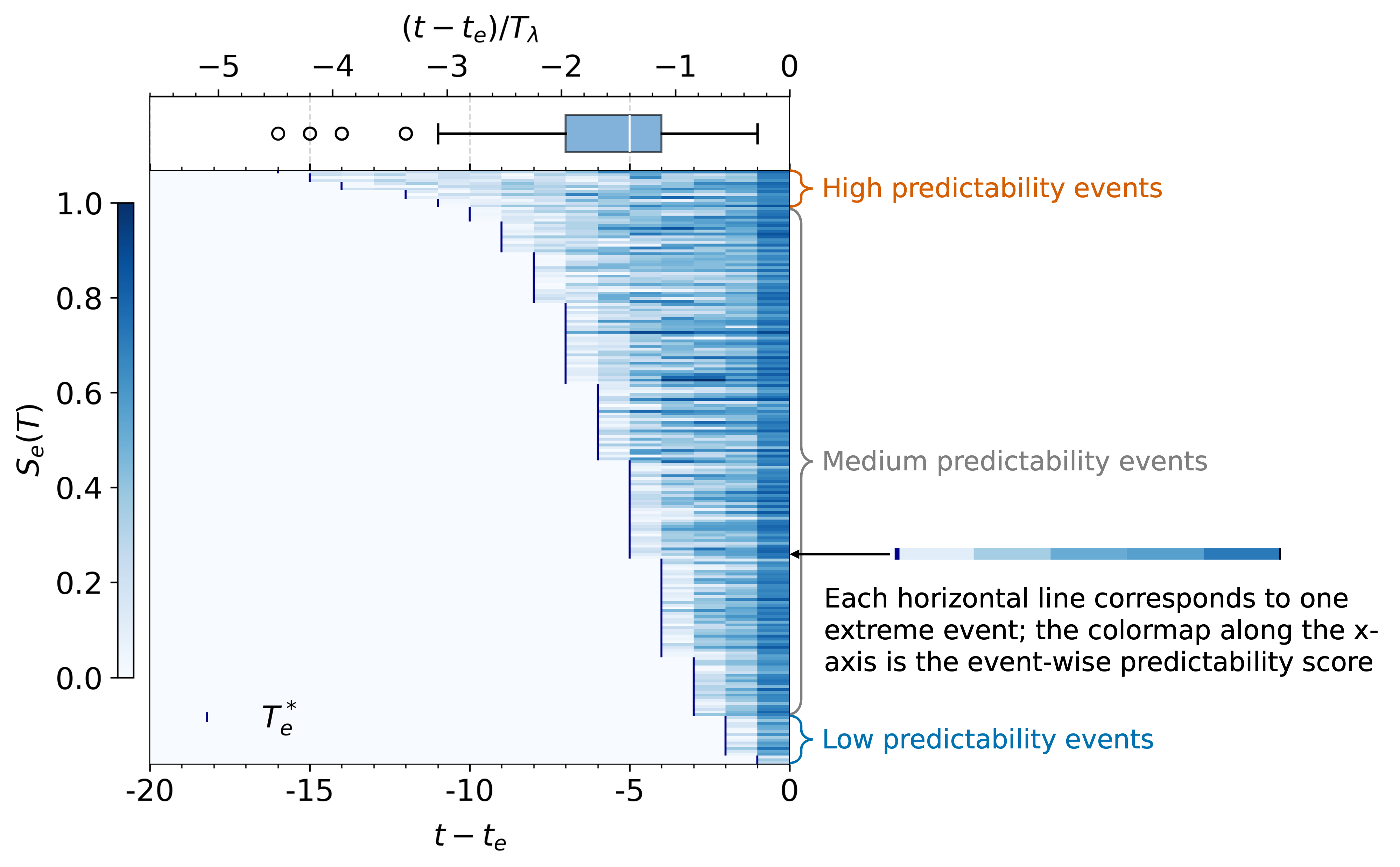}
\caption{\label{fig:1} Predictability evolution of extreme events prior to occurrence. The heatmap shows ${S}_e(T)$ against $t-t_e$ (bottom axis) and normalized time $(t-t_e)/T_{\lambda}$(top axis), with darker shadings indicating higher predictability. 
Vertical ticks to the left of each shading mark the individual predictability limit (horizons) $T^{*}_e$ for each corresponding extreme event. 
The box plot above the heatmap summarizes the distribution of the event-wise predictability score ($S_{e}(T)$) for all events.
The right axis indicates the predictability regime of each event (high predictability, medium predictability, or low predictability) based on its corresponding $T^*_e$. }
\end{figure}

Fig.~\ref{fig:1} displays ${S}_e(T)$ (blue colormap) along with their associated predictability limits $T^*_e$ for each extreme event in the test set (i.e., $N_{events}=211$), where we also categorized extreme events into three regimes (right axis): a high-predictability regime ($T^*_e>3T_\lambda$), a low-predictability regime (bottom $8\%$ quantile, chosen to approximately match the number of events in the high-predictability regime), and an intermediate regime comprising the remaining events.
The horizontal axis at the bottom shows the time relative to the event, $t-t_e \le 0$ in simulation time units, while the top axis shows the simulation time units normalized by the Lyapunov time $T_\lambda$.
Note that increasingly negative values of $t-t_e$ correspond to longer lead times, i.e., earlier predictions.
The figure demonstrates that predictability rapidly declines with increasing lead time.
Most extreme events lose predictability within approximately 7 simulation time units, corresponding to 1-2 Lyapunov times ($T_{\lambda}$). 
This is visible in the heatmap as the colors transition from dark blue (high ${S}_e(T)$) on the right (short lead time) to light blue (low ${S}_e(T)$) on the left (long lead time).
However, certain extreme events exhibit remarkably long predictability limits, extending up to 15 simulation time units, corresponding to more than 4 Lyapunov times.

Given this event-dependent nature of predictability, that we were able to detect thanks to the proposed framework, we ask ourselves: \textit{what is driving different predictability horizons for different events?}
To answer this question, we first attempt to understand whether the predictability of extreme events in the Kolmogorov flow is governed primarily by large-scale coherent structures. 
To this end, we repeat the forecasts obtained with the diffusion-based autoregressive probabilistic model after systematically removing small-scale components from the initial conditions. 
More specifically, we apply an isotropic spectral cutoff in Fourier space, where modes with radial wavenumber $\smash k=(k_x^2+k_y^2)^{1/2}>k_c$ are removed. 
For our computational domain ($L=2\pi$, $N=128$), the cutoff is defined as $k_c = r_c k_{\max}$, with $r_c$ being the cutoff ratio and $k_{\max}=N/2=64$. 
By progressively decreasing $r_c$, we can determine the smallest spatial scale whose removal does not alter the event-wise predictability limits. 

\begin{figure}[H]
\centering
\includegraphics[width=0.95\textwidth]{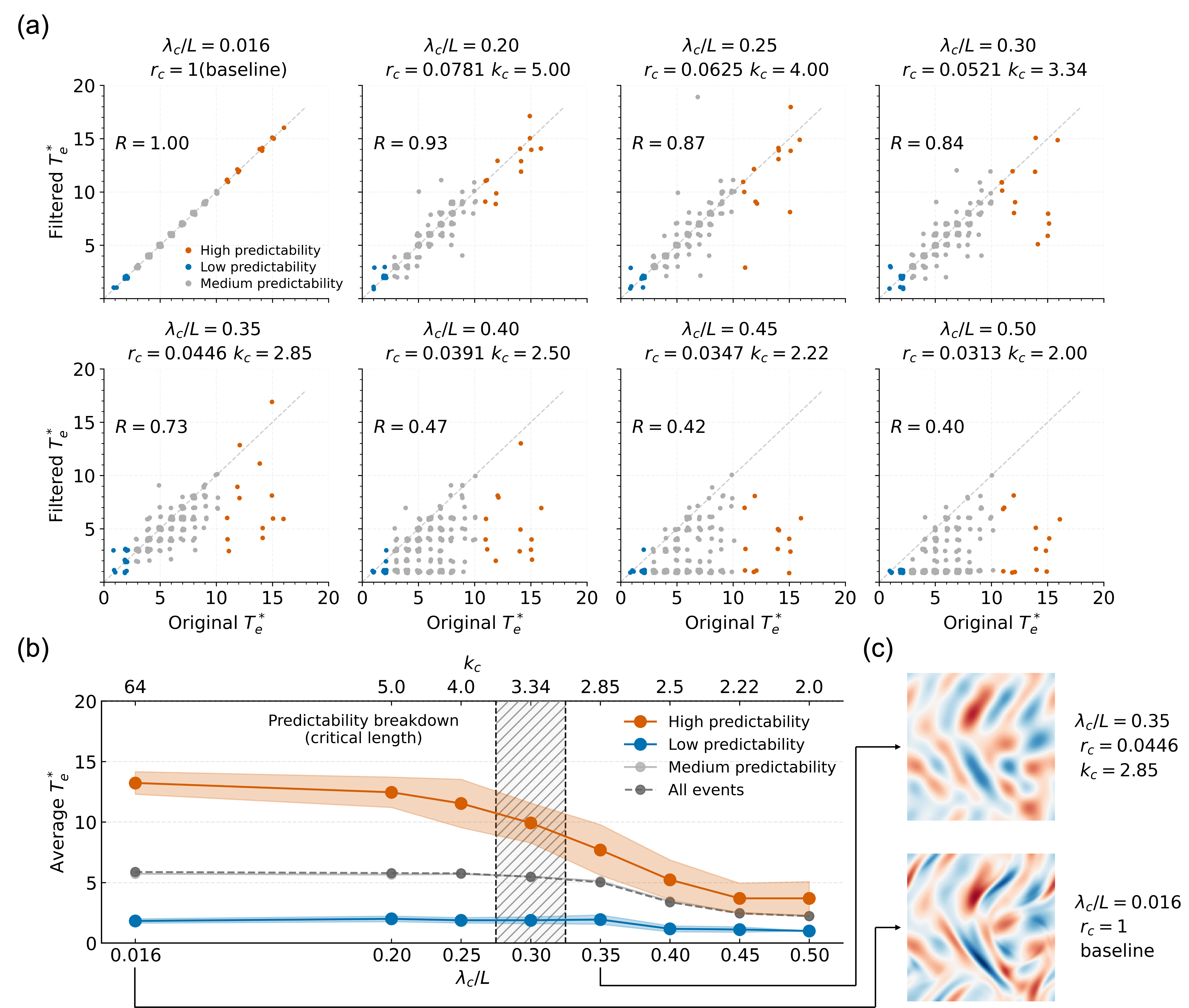}
\caption{Influence of spatial filtering scale $r_c$ on predictability limit $T^*_e$. (a) Scatter plots of filtered versus baseline $T^*_e$. (b) Evolution of $T^*_e$ for high (orange) and low (blue) predictability limits events. The vertical dashed area ($\lambda_c/L = 0.3$) marks the region of predictability breakdown. (c) Visualization of structural degradation in vorticity fields, with the baseline presented on the right and the breakdown scale ($\lambda_c/L=0.35$) on the left.}
\label{fig:2} 
\end{figure}

Fig.~\ref{fig:2} illustrates how predictability changes when applying the filtering just introduced, whereby the selected physical wavelengths $\lambda_c = 2\pi/k_c$ and corresponding cutoff ratios are listed on each subfigure (these values are also reported in a table in the Supplemental Material for the reader's convenience). 
The physical wavelength $\lambda_c/L$ is increased linearly, corresponding to a decrease in the cutoff ratio $r_c$, thereby progressively removing larger spatial structures.
Fig.~\ref{fig:2}(a) depicts the scatter plots of filtered versus original (i.e., without filtering) predictability limits $T_e^*$. 
We observe that the predictability limits after filtering remain similar to the original ones for low-predictability (blue dots) and medium-predictability (grey dots) extreme events up to a physical wavelength $\lambda_c = 0.30L$ (the points representing extreme events in the scatter plots remain clustered along the diagonal). 
For high-predictability extreme events (orange dots), we start to see a breakdown in predictability slightly earlier, at physical wavelengths between $\lambda_c = 0.25L$ and $\lambda_c = 0.30L$.
As a whole, extreme events are  insensitive to small-scale filtering up to $\lambda_c = 0.30L$, as shown by the relatively high correlation coefficients ($R \geq 0.84$).
When removing larger scales corresponding to wavelength $\lambda_c = 0.30L$, we start seeing a breakdown in predictability across all extreme events (i.e., high-, medium- and low-predictability events), as shown by the significantly degraded correlation coefficient (($R \leq 0.73$) that is reflected by the spread of points off the diagonal. 

To further assess this behavior, in Fig.~\ref{fig:2}(b), we show the evolution of the average predictability limits $\langle T^*_e \rangle$ for each extreme event category (high-, medium-, low-predictability) as a function of the physical wavelength filter $\lambda_c/L$ applied. 
High-predictability events (orange line) retain long forecast horizons (limits) until structures with physical wavelengths between $\lambda_c = 0.25L$ and $\lambda_c = 0.30L$ are removed, consistent with the results shown in Fig.~\ref{fig:2}(a). 
Once structures with a physical wavelength $\lambda_c = 0.30L$ are removed, we start seeing a breakdown in predictability across all events, namely high-, medium-, and low-predictability.
This critical scale is identified by comparing the distribution of predictability limits at each filtering scale with the baseline (unfiltered) distribution using the Wasserstein distance, with statistical significance assessed via a paired permutation test (see Supplemental Material).

\begin{figure}[H]
\centering
\includegraphics[width=1.\textwidth]{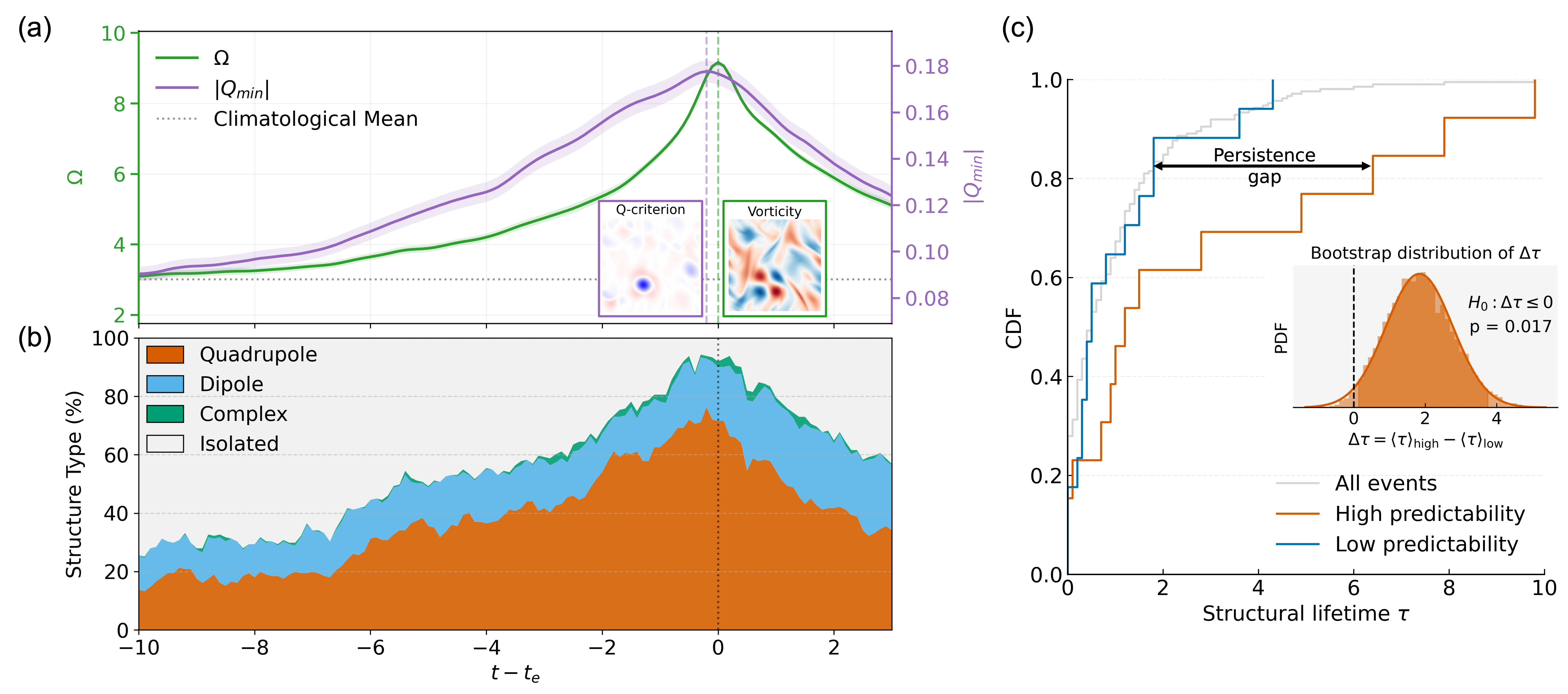}
\caption{(a) Ensemble-averaged evolution of $\Omega$ (green, left axis) and $|Q_{min}|$ (purple, right axis), with shaded regions denoting one standard deviation. The insets display representative snapshots of the $Q$-criterion and vorticity at the peak of the extreme event. (b) Temporal evolution of coherent-structure statistics within the analysis window centered on the strain core, defined by the maximum of $|Q_{min}|$. (c) Cumulative distribution functions (CDFs) of structural lifetimes $\tau$ for high-predictability (orange) and low-predictability (blue) events. The inset shows the bootstrap distribution of the mean lifetime difference $\Delta\tau = \langle\tau\rangle_{\mathrm{high}} - \langle\tau\rangle_{\mathrm{low}}$, yielding $p = 0.017$.}
\label{fig:3} 
\end{figure}

We further note that the breakdown wavelength identified through statistical analysis corresponds to a degradation of structural coherence in the flow, as illustrated in Fig.~\ref{fig:3}(c).
Specifically, we compare the flow field for two filtering scales, namely $\lambda_c = 0.016L$ (i.e., no filtering) and $\lambda_c = 0.35L$. 
The baseline flow exhibits sharp coherent structures, whereas strong filtering retains only smooth vortex clusters.
In particular, the removal of structures with characteristic wavelengths of order $\lambda_c/L \approx 0.3$ eliminates the distinct flow organization, leading to a rapid loss of predictability.
The negligible sensitivity of predictability to the removal of small-scale structures shown in Fig.~\ref{fig:2} can be understood in terms of the energetic hierarchy of two-dimensional turbulence~\cite{kraichnan1971inertial}. 
While small-scale motions contribute substantially to enstrophy, they carry only a minor fraction of the total kinetic energy and are therefore less influential in shaping the large-scale flow evolution~\cite{boffetta2012two}. 
As a result, filtering out high-wavenumber modes primarily removes fine-scale fluctuations without substantially altering the dominant coherent structures~\cite{farge1992wavelet}.
Predictability is therefore constrained by the evolution of energy-dominant large-scale structures, which persist over longer timescales and govern the growth of forecast errors.

These observations naturally raise the question of which coherent structures are responsible for extreme events, and which of their characteristics control the predictability limit.
We therefore examine the flow structure associated with extreme events, and their relation to predictability.

Fig.~\ref{fig:3}(a) presents the composite temporal evolution of enstrophy (green line) and the local minimum of the $Q$-criterion (purple line), conditionally averaged over all detected extreme events and aligned at the event time.
Both quantities exhibit pronounced variations near the extreme, with the intensification of negative $Q$ (represented as $|Q_{min}|$ in purple curve) systematically preceding the peak in enstrophy (green curve).
This temporal offset reveals a strong correlation between intense strain-dominated regions and subsequent enstrophy amplification, suggesting that the emergence of a strain core acts as a precursor to enstrophy bursts~\cite{farazmand2019extreme}.
As shown in the inset snapshots, while the vorticity field is spatially complex and fragmented, the strain core identified by the minimum of $Q$ provides a clear and localized reference point around which dynamically relevant coherent structures organize.
This makes the strain core a more effective anchor than vorticity extrema for identifying the flow structures that govern extreme events and influence their predictability.

Motivated by this observation, we apply a structure classifier to the flow field within a local window of size $\lambda_c/L \approx 0.3$ centered on the strain core. 
This wavelength corresponds to the statistically identified critical length at which predictability begins to break down (Fig.~\ref{fig:2}).
Based on the number of dynamically significant vortex cores contained within this window, identified after applying an intensity threshold to exclude weak vortices, the local flow configuration is classified into four categories: isolated (0–1 vortex), dipole (2 vortices), quadrupole (3–4 vortices), and complex configurations (more than 4 vortices). The 3–4 category is labeled as quadrupole because quadrupolar arrangements may occasionally appear with only three detected cores when one vortex falls below the intensity threshold or lies partially outside the detection window.

Fig.~\ref{fig:3}(b) shows the temporal evolution of the relative occurrence of these structure types, demonstrating that quadrupoles become increasingly dominant as the system approaches the enstrophy peak.
This trend suggests that quadrupoles constitute a characteristic dynamical configuration underlying extreme events, motivating a closer examination of their temporal persistence and their impact on predictability.
We therefore quantify the structural lifetime of quadrupolar configurations in the vicinity of the strain core and examine its relationship to event predictability.
The structural lifetime $\tau$ is defined as the continuous time interval over which a quadrupolar structure persists prior to the enstrophy maximum in a Lagrangian frame; events whose peak configuration is not classified as a quadrupole are assigned a lifetime of zero.
The same intensity threshold and window size as in Fig.\ref{fig:3}(b) are employed to ensure consistency of the structural classification.

As demonstrated in Fig.~\ref{fig:3}(c), a pronounced separation in structural persistence emerges between high-predictability (orange) and low-predictability (blue) events.
The cumulative distribution functions (CDFs) show that high-predictability events are systematically associated with longer-lived quadrupoles.
This separation is further supported by a one-sided bootstrap test of the lifetime difference, $\Delta\tau=\langle\tau\rangle_{\mathrm{high}}-\langle\tau\rangle_{\mathrm{low}}$ (inset), testing the hypothesis $\Delta\tau<0$, with $p=0.017$ based on 20,000 bootstrap resamples.

These results indicate that the temporal stability of quadrupoles plays an important role in shaping the predictability of extreme events in Kolmogorov flow.
The persistence of the quadrupole suggests a structural resilience against the turbulent background, which limits local error growth compared to low-predictability events~\cite{boffetta2017chaos}. 
Consequently, the flow evolution with long-lived quadrupoles remains comparatively more predictable, explaining the observed persistence gap.
Notably, this enhanced predictability is not associated with the largest instantaneous enstrophy values. 
Instead, the longest predictable extreme events typically exhibit moderate enstrophy peaks accompanied by long-lived coherent structures. 
By contrast, the least predictable events tend to coincide with the strongest enstrophy bursts accompanied by short-lived coherent structures, reflecting a general loss of structural coherence and predictability.

We further verified that the distribution of predictability limits are robust across different diffusion architectures.
We tested three denoising architectures, including a U-Net with ConvNeXt blocks, a U-Net with ConvNeXt blocks and attention, and a U-Net with ResNet blocks and attention -- all yield nearly identical statistics of $T^*_e$ (see Supplemental Material). 
These results demonstrate that the observed predictability limits and event-wise hierarchies are governed primarily by the underlying flow dynamics rather than by model-specific biases.

To quantitatively validate the predictability estimates obtained with the diffusion models, we perform DNS ensemble simulations via perturbations of the initial conditions for a subset of extreme events, initializing 2 $T_{\lambda}$ prior to the events' peak.
Despite the fundamentally different constructions of the two approaches, we find that the diffusion-based predictability estimates qualitatively align with the divergence behavior observed in DNS ensembles (see Supplemental Material).

\section{Conclusions}\label{sec:conclusions}
In this work, we introduced a diffusion-based ensemble framework to quantify predictability in complex systems from a data-driven, event-wise perspective. 
Rather than relying on predefined perturbation ensembles or explicit assumptions about the underlying dynamics, the framework provides a statistical characterization of forecast uncertainty directly from data, enabling scalable and reproducible predictability analyses.

Applied to the two-dimensional turbulent Kolmogorov flow, our results reveal a pronounced heterogeneity in the predictability of extreme events, with predictability limits spanning from approximately 1 to more than 4 Lyapunov times. 
Crucially, this variability cannot be explained by event intensity alone. 
Instead, highly predictable extremes are consistently associated with persistent, large-scale coherent structures, whereas the least predictable events coincide with intense, small-scale-dominated bursts and a loss of structural coherence.
Through scale-filtering and structural analyses, we demonstrate that predictability is ultimately constrained by the evolution and persistence of energy-dominant large-scale flow structures. 
Long-lived quadrupolar configurations emerge as a representative example of such structures, providing a concrete physical link between flow organization and extended predictability horizons.

More broadly, these findings reveal that the predictability of extreme events in turbulent flows is fundamentally constrained by the persistence of coherent structures that organize the flow dynamics. 
By linking event-wise predictability to the structural stability of quadrupoles, the proposed framework provides a route to uncover predictability hierarchies of extreme events and their structural origins in high-dimensional chaotic systems, with implications for forecasting, uncertainty quantification, and extreme-event analysis.


\clearpage


\clearpage
\bibliographystyle{unsrt}  
\bibliography{references}  


\end{document}